\documentclass{article}
\usepackage{epsfig}
\begin{document}

\title{\bf Estimating the inelasticity with the information theory
approach} 
\author{F.S.Navarra$^{1}$\thanks{e-mail: NAVARRA@IF.USP.BR},
O.V.Utyuzh$^{2}$\thanks{e-mail: utyuzh@fuw.edu.pl},
G.Wilk$^{2}$\thanks{e-mail: wilk@fuw.edu.pl} and Z.W\l odarczyk$^{3}$
\thanks{e-mail: wlod@pu.kielce.pl}\\[2ex]  
$^{1}${\it Instituto de F\'{\i}sica, Universidade de S\~{a}o Paulo,
        S\~{a}o Paulo, SP, Brazil}\\[1ex]
$^{2}${\it The Andrzej So\l tan Institute for Nuclear Studies,
           Warsaw, Poland}\\[1ex]
$^{3}${\it Institute of Physics, \'Swi\c{e}tokrzyska Academy,
           Kielce, Poland}}
\date{\today}
\maketitle

\begin{abstract}
Using the  information theory approach, in both  its extensive and
nonextensive versions, we estimate the inelasticity parameter $K$ of
hadronic 
reactions together with  its distribution and energy dependence from 
$p\bar{p}$ and $pp$ data. We find that the inelasticity remains
essentially constant in energy except for a variation around $K\sim 0.5$,
as was originally expected.\\      

\end{abstract}

\section{Introduction}
The inelasticity $K$ ($0<K<1$) of a reaction has well established 
importance  in working with data from cosmic ray cascades
\cite{SWWW} (cf. also \cite{CAS1,CAS2,CAS3} and references therein).
It tells us what fraction of the energy of a projectile is used for
production of secondaries and what fraction flows further along the
cascade chain. In cosmic ray observables, $K$ in fact appears 
in some combination which also contains the mean free path for the particle
propagation in the  atmosphere, or equivalently, the total
inelastic cross section $\sigma_{in}$. This makes $K$  difficult
to estimate because of the freedom available to attribute the observed
effects either to $K(s)$ or to $\sigma(s)$. It was therefore proposed
in \cite{SWWW} that in order  to extract $K(s)$ unambiguously  from cosmic ray
experiments one should  analyse {\it simultaneously} the data from  (at
least) two different types of experiments for  which combinations of $K$
and $\sigma_{in}$ are different \cite{FOOT1}.\\

Nowadays there is a strong tendency to replace  inelasticity  and
simple energy flow models by  more refined and complicated models of
multiparticle production (see \cite{Wibig,EW,E} and references
therein). In their present formulations such models differ
substantially among themselves, concerning both  their physical 
basis and  the (usually very large) number  of parameters used, and
lead to quite different, sometimes even contradictory, predictions
\cite{EW,E}. While developing models is necessary for the global
understanding of cosmic ray physics, for the purpose of studying
energy flow it may be desirable to have a more economical description
of high energy collisions, involving only a small number of
parameters. This is one of the advantages of working with the concept
of inelasticity \cite{SWWW,LEADING}.\\       

The inelasticity is also a very important quantity in phenomenological
descriptions of hadronic and nuclear collisions in terms of
statistical models of multiparticle production processes
\cite{INEL,Fernando}. In this case it enters either explicitly, as a
single parameter $K$ defining  the initial energy $M=K\cdot
\sqrt{s}$ to be  hadronized, or implicitly, when one finds that
out of numerous parameters of the model only the combination
leading to the given fraction of available energy to be converted into
produced secondaries, $K\cdot\sqrt{s}$, is important, see
\cite{Chao}. Attempts to  estimate it are thus fully 
justifiable. In \cite{INEL,Fernando} the calculations of the  inelasticity
distribution, $\chi(K)$ (and its energy dependence), based on the
assumed dominance of high energy multiparticle processes by gluonic 
interactions, was presented. In the other calculations the mean 
inelasticity $K$
and its possible energy dependence was simply estimated either by
using  thermal-like model formulas applied to collider $\bar{p}p$
data \cite{UA5,Tevatron,P238,UA7,DeltaN} (like, for example, in
\cite{Chou}) or by some other means \cite{Inel}.\\   

In this paper we address this problem again, this time by means of
the information theory approach both in its extensive \cite{MaxEnt} and
nonextensive \cite{T,WWq} versions. The idea is to describe the
available data by using only a {\it truly minimal amount of
information} avoiding therefore any unfounded and unnecessary
assumptions. This is done by attributing to the measured distributions
(written in terms of the suitable probability distributions) an 
information entropy $S$ and maximizing it subject to constraints which 
account for our {\it a
priori} knowledge of the process under consideration. As a result one 
gets the {\it most probable} and {\it least biased} distribution
describing these data, which is not influenced by anything else
besides the available information. In such approach the inelasticity
$K$ emerges as the only real parameter; all other quantities being well
defined functions of it. We attempt to clarify here the role of the 
inelasticity by using both the extensive and nonextensive versions of 
information theory. All necessary background  on the information
theory approach needed in the present context is given in the next
Section. Section 3 contains our results for the $p\bar{p}$ and $pp$
collisions. Our conclusions and summary are presented in the last Section.\\

\section{General ideas of information theory approach}

As presented at length in \cite{MaxEnt} (where further details and
references can be found) the  information theory approach provides us, by
definition, with the {\it most probable, least biased} estimation of
a   probability distribution $\{ p_i, i=1,\dots,n\}$ using only
knowledge of a finite number of observables $\{ F_k, k=1,\dots, n\}$
of some physical quantities obtained by means of $p_i$ and defined
as:  
\begin{equation}
F_k = \langle {F}_k\rangle = \sum_{i=1}^{n} p_i\, {F}_k^{(i)}
.\label{eq:average} 
\end{equation}
One is looking for such $\{ p_i\}$ which contain only information
provided by $\{F_k\}$ and nothing more, i.e., which contain {\it
minimal} information. The information connected with $\{ p_i\}$ is
quantified by the Shannon information entropy defined as:
\begin{equation}
S\, =\, -\Sigma_i p_i\ln p_i .\label{eq:Shannon}
\end{equation}
Minimum information corresponds to maximum entropy $S$, therefore the
$\{p_i\}$ we are looking for are obtained by maximizing the
information entropy $S$ under conditions imposed by the measured
observables $\{ F_k\}$ as given by eq.(\ref{eq:average}). They result
in a set of Lagrange multipliers $\{ \lambda_k, k=1,\dots,r\}$ and
the generic form of $\{p_i\}$ we are looking for is \cite{MaxEnt}: 
\begin{equation}
p_i = \frac{1}{Z}\, \exp\left[ - \sum_{k=1}^r\, \lambda_k \cdot
F_k^{(i)} \right], \label{eq:pi}
\end{equation}
where $Z$ is obtained from the normalization condition
$\sum_{i=1}^np_i = 1$.\\

Such approach was applied long time ago to experimental data on
multiparticle production with the aim at establishing the minimum amount
of information needed to describe them \cite{Chao}. The rationale was
to understand what makes all the apparently disparate (if not outright
contradictory) models of that period  fit (equally well) the
data. The result was striking and very instructive \cite{Chao}: the  data
considered (multiplicity and momentum distributions) contained only
very limited amount of information, which could be expressed in the
form of the following two observations: $(i)$ the available phase
space in which  particles are produced  is limited (i.e., there
is some sort of $p_T$ cut-off) and  $(ii)$ only a part $K\in (0,1)$
of the available energy $\sqrt{s}$ is used to produce the observed 
secondaries, the rest being taken away by the so called leading
particles (i.e., inelasticity emerges as one of the cornerstone
characteristics of reaction). All other assumptions, different for
different models (based on different, sometimes even contradictory,
physical pictures of the collision process) were therefore spurious
and as such they could be safely dropped out without spoiling the
agreement with experimental data. In fact, closer scrutiny of these
models showed that they all contained, explicitly or implicitly,
precisely those two assumptions mentioned above and that was the true
reason of their agreement with data.\\

In this paper we are therefore following the same line of approach
with the aim at deducing from the available data for  the inelasticity
parameter $K$. Notice that the formula (\ref{eq:pi}) resembles the
statistical model formulas based on the Boltzmann-Gibbs statistics as 
used in \cite{Chou}. However, in (\ref{eq:pi}) no thermal
equilibrium is assumed and all $\lambda_k$ (i.e., among others also the 
"partition temperature" $T$ in \cite{Chou}) are given by the
corresponding constraint equation (\ref{eq:average}) whereas
normalization fixes $Z$, which is a free parameter in \cite{Chou}.\\

This approach can be generalized to systems which cannot be described
by Boltzmann-Gibbs (BG) statistics because of either the existence of
some sort of long-range correlations (or memory) effects or fractal
structure of their phase space or because of the existence of 
some intrinsic fluctuations in the system under consideration. It
turns out that such systems are nonextensive and therefore must be
described by a nonextensive generalization of BG statistics, for
example by the so called Tsallis statistics \cite{T} defined by the
following form of the entropy:
\begin{equation}
S_q\, =\, - \frac{1}{1-q}\Sigma_i\left(1 - p_i^q\right)
.\label{eq:Tsallis} 
\end{equation}
It is characterized by the  nonextensivity parameter $q$ such that for two
independent systems $A$ and $B$
\begin{equation}
S_{q(A+B)} = S_{qA} + S_{qB} + (1-q)S_{qA}S_{qB}. \label{eq:SAB}
\end{equation}
Notice that in the limit $q\rightarrow 1$ one recovers the previous
form of Boltzmann-Gibbs-Shannon entropy (\ref{eq:Shannon}).
Maximazing $S_q$ under constraints, which are now given in the 
form \cite{FOOT2}:
\begin{equation}
F^{(q)}_k = \langle {F}_k\rangle_q = \sum_{i=1}^{n} [p_i]^q\,
{F}^{(q,i)}_k ,\label{eq:qaverage} 
\end{equation}
results in the following power-like form of the (most probable, least
biased) probability distribution: 
\begin{equation}
p_i\, =\, p^{(q)}_{i}\, =\, \frac{1}{Z_q}\, \exp_q\left[ -
\sum_{k=1}^r\, \lambda_k \cdot F^{(q,i)}_k \right], \label{eq:qpi}
\end{equation}
where $Z_q$ is obtained from the normalization condition
$\sum_{i=1}^np^{(q)}_i = 1$. 
and where
\begin{equation}
\exp_q \left(- \frac{x}{\Lambda}\right) \stackrel{\it def}{=} \left[1
- (1-q) \left(\frac{x}{\Lambda}\right)\right]^{\frac{1}{1-q}}
.\label{eq:qexp}   
\end{equation}
Of special interest to us here will be the fact that intrinsic
fluctuations in the system, represented by fluctuations in the
parameter $1/\Lambda$ in the exponential distribution of the form
$\sim \exp(-x/\Lambda)$ result in its nonextensivity with parameter
$q$ given by a normalized variation of fluctuation of the  parameter
$1/\Lambda$ \cite{WWq}:
\begin{equation}
q = 1 \pm \frac{\left\langle
\left(\frac{1}{\Lambda}\right)^2\right\rangle - \left\langle 
\frac{1}{\Lambda}\right\rangle^2}{\left\langle
\frac{1}{\Lambda}\right\rangle^2}. \label{eq:qfluct} 
\end{equation}
So far this has been proven only for fluctuations of $1/\Lambda$
given in the form of the gamma distribution, 
\begin{equation}
f\left(\frac{1}{\Lambda}\right) = \frac{\mu}{\Gamma(\alpha)}
\left(\frac{\mu}{\Lambda}\right)^{\alpha - 1} 
           \exp \left(-\frac{\mu}{\Lambda}\right), \label{eq:fluct}
\end{equation}
but this conjecture seems to be valid also in general \cite{CBeck}           .\\

\section{Inelasticity obtained from analysis of the $\bar{p}p$
collider and $pp$ fixed target data} 

The available information in this case consists of: 
\begin{itemize}
\item[$(i)$] The mean multiplicity of charged secondaries, 
$\langle n_{ch}\rangle$, produced in nonsingle diffractive reactions
at given energy $\sqrt{s}$, which can be parametrized as $\langle
n_{ch}\rangle = A + B\ln s + C \ln^2 s$ or as $\langle n_{ch}\rangle
= D + E s^{\gamma}$ (see  \cite{N}). In what follows we shall assume,
for simplicity, that they are pions with mass $\mu = 0.14$ GeV. Out
of it, we shall construct and use the total mean number of produced
particles assuming it to be $N= \frac{3}{2}\langle n_{ch}(s)\rangle$.  
\item[$(ii)$] The observation that the phase space filled by the
produced secondaries is essentially one-dimensional with limited (and
only slowly growing with energy) transverse momenta \cite{NB}:
$\langle p_T\rangle = 0.3 + 0.044\ln
\left(\frac{\sqrt{s}}{20}\right)$ \cite{FOOTA}.  
\item[$(iii)$] The rapidity distributions of charged secondaries,
$dN(s)/dy=\int d^2p_T \frac{dN}{d^3p}$, provided either by collider
experiments \cite{UA5,Tevatron,P238,UA7} or by the fixed target
experiment \cite{NucData2}. They are available only in a limited
range of the rapidity space, depending on the details of the
experimental setup \cite{NB}.    
\end{itemize}
Following the results obtained in \cite{Chao} we expect (and
therefore assume in what follows) that only a part $W=K\sqrt{s}$ of
the total energy $\sqrt{s}$ is   used to produce 
secondaries in the central region of the  investigated reaction. The
inelasticity $K=K(s)$ will therefore be the main quantity we shall
investigate.\\ 

We start with the  information theory approach in its extensive version.
In this case the relevant probability distribution defining 
information entropy (\ref{eq:Shannon}) is given by
\begin{equation}
p(y) = p(y;N,W=K\sqrt{s}) = \frac{1}{N}\cdot \frac{dN}{dy}
\label{eq:p} 
\end{equation}
whereas constraint (\ref{eq:average}) is just the energy conservation
(here $\mu_T= \sqrt{\mu^2 + \langle p_T\rangle^2}$ and $\langle
E\rangle$ is the  mean energy per produced particle) \cite{OMT}:
\begin{equation}
\int_{-Y_m}^{Y_m}\, dy\, \left[ \mu_T\, \cosh y\right]\, p(y) =
\frac{W}{N} = \langle E\rangle = \frac{K\cdot \sqrt{s}}{N} .  \label{eq:W}
\end{equation}
The limits of the relevant longitudinal phase space are
\begin{equation}
Y_m = \ln \left\{ \frac{W'}{2\mu_T}
       \left[1 + \left( 1 -
\frac{4\mu^2_T}{W'^2}\right)^{\frac{1}{2}}\right] \right\},\qquad 
W' = W - (N-2)\mu_T . \label{eq:Ymax} 
\end{equation}
We would like to stress that throughout this paper the central region
of the reaction, i.e., the region populated by produced particles
distributed according to $p(y)$ (or $p_q(y)$ later on), is always
defined by eq. ({\ref{eq:Ymax}).  Therefore $y\in \left(-Y_m,Y_m\right)$
in the CMS and we do not choose arbitrary cuts in rapidity space
(as, for example, in \cite{Chao}). Following now the steps
mentioned in Section 2, i.e., minimazing the respective information
entropy (\ref{eq:Shannon}) with the  constraint given by (\ref{eq:W}), we
arrive at \cite{MaxEnt}: 
\begin{equation}
p(y) = \frac{1}{N}\frac{dN(y)}{dy} = \frac{1}{Z}\, \exp\left(- \beta
\mu_T \cosh y\right) \label{eq:formula}
\end{equation}
with $\beta = \beta(W,N,\mu_T)$ given by solving eq. (\ref{eq:W}) and
\begin{equation}
Z = Z(W,N,\mu_t) = \int^{Y_m}_{-Y_m}\, dy \exp\left(- \beta \mu_T \cosh
y\right) \label{eq:Z}
\end{equation}
given by the normalization condition, $\int dy p(y) =1$. Notice that
in such an approach the "inverse temperature" $\beta$ and the
normalization $Z$ depend on our input information, i.e., on $N$ and
$\mu_T$, and on the assumed inelasticity $K$ (via $W=K\cdot
\sqrt{s}$), which is our free parameter. They are therefore maximally
correlated which means that the shape of distribution $p(y)$ (given by
$\beta$) and its height (given by $Z$) are not independent of each
other. This is in sharp contrast to the  approaches presented before in
\cite{Chou} where {\it both} $\beta$ (called "partition
temperature") and the normalization (our $Z$) were treated as two
independent parameters. Because of the symmetry of the  colliding system
momentum conservation does not impose any additional constraint
here and we are left with  $\beta$ being the only Lagrange
multiplier to be calculated from  eq.(\ref{eq:W}) for each energy
$W$ and multiplicity $N$.\\

Notice that (\ref{eq:formula}), although formally resembling formulae
obtained in thermal models \cite{Bec}, has a much wider range of
applicability as it is not connected with any assumption of thermal 
equilibrium. Actually it can be written in the scaling-like
form: 
\begin{equation}
p(y)\, =\, \frac{1}{Z}\exp\left[ - \bar{\beta}\cdot \frac{\mu_T \cosh
y}{\langle E\rangle}\right], \label{eq:formulas}
\end{equation}
where $\langle E\rangle = W/N$ is the mean energy per produced
particle and $\bar{\beta} = \beta \cdot W/N$. Plotting $\bar{\beta}$
as a function of $\langle E\rangle$ one observes that for the
minimal number of produced secondaries ($N\rightarrow 2$)
$\bar{\beta} \rightarrow -\infty$, whereas for the maximal number
($N\rightarrow N_{max} = W/\mu_T$) $\bar{\beta} \rightarrow +\infty$.
There is also an intermediate region in which $\bar{\beta}$ remains fairly
constant leading to an approximate "plateau" in $p(y)$. In this
region "partition temperature" $1/\beta$ and inelasticity $K$ are
related in a very simple way, namely
\begin{equation}
\beta \, \simeq \, \frac{N}{K\cdot \sqrt{s}} . \label{eq:betaK}
\end{equation}
Actually, $\beta > 0$ only for $N > N_0 = 2\ln \left(N_{max}\right)$
and "plateau" occurs only for $N \simeq N_0$. It clearly shows that
$T=1/\beta$, called sometimes "partition temperature" \cite{Chou},
is a measure of energy available per produced particle (which
therefore depends also on inelasticity).\\ 

\begin{figure}[h]
\begin{minipage}[t]{0.475\linewidth}
\centering
\includegraphics[height=7.0cm,width=7.0cm]{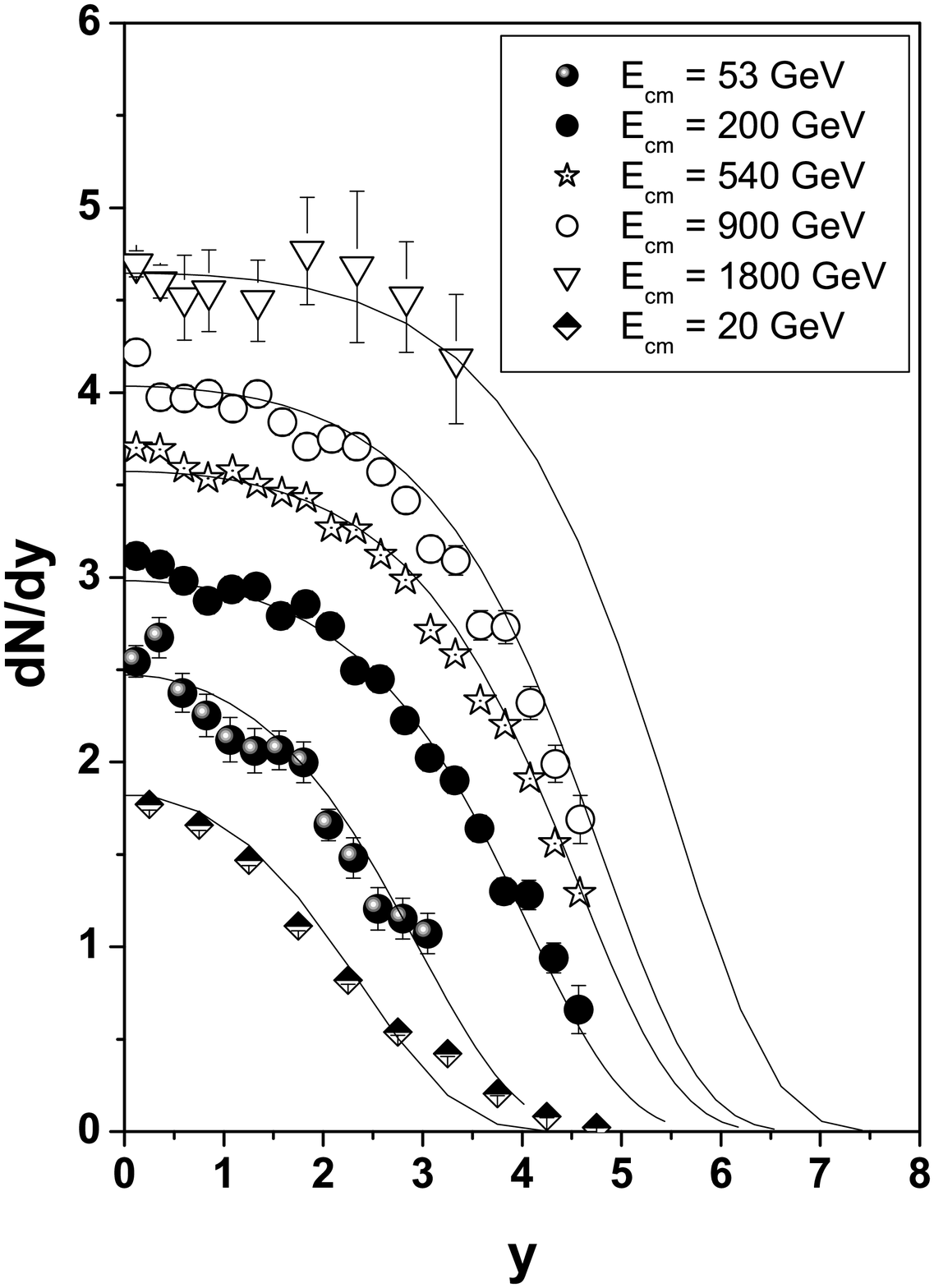}
\caption{Rapidity spectra obtained by UA5 \cite{UA5} and Tevatron
\cite{Tevatron} fitted by formula (\ref{eq:formula}). For
completeness results for $pp$ data at $20$ GeV \cite{NucData2} are
also shown. The corresponding values of inelasticity $K$ are listed
in Table \ref{table:TI}.} 
\label{fig:fig1a}
\end{minipage}\hspace{2mm}
\begin{minipage}[t]{0.475\linewidth}
\centering
\includegraphics[height=7.cm,width=7.cm]{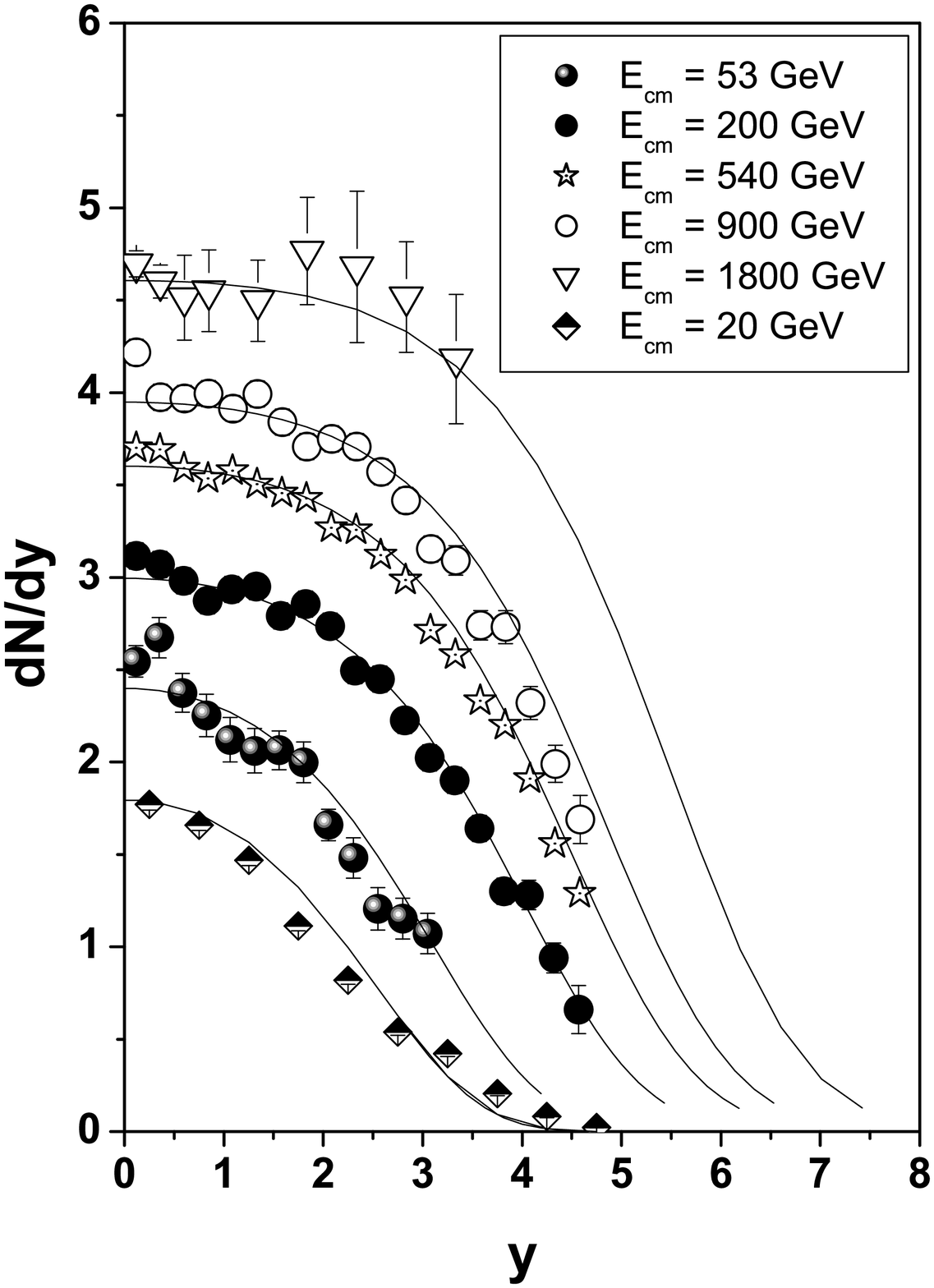}
\caption{Rapidity spectra obtained by UA5 \cite{UA5} and Tevatron
\cite{Tevatron} fitted by formula (\ref{eq:formulaq}). For
completeness results for $pp$ data at $20$ GeV \cite{NucData2} are
also shown. The corresponding values of $q-$inelasticity $\kappa_q$
and nonextensivity parameter $q$ are listed in Table \ref{table:TI}.} 
\label{fig:fig1b}
\end{minipage}
\end{figure}

In the case of the nonextensive version of information entropy the 
energy conservation constraint is given by
\begin{equation}
\int_{-Y_m}^{Y_m}\, dy\, \left[ \mu_T\, \cosh y\right]\, [p_q(y)]^q =
\frac{W_q}{N} = \frac{\kappa_q\cdot \sqrt{s}}{N} .  \label{eq:Wq}
\end{equation}
and maximization of the corresponding Tsallis entropy
(\ref{eq:Tsallis}) results in 
\begin{equation}
p_q(y) = \frac{1}{N}\frac{dN_q}{dy} = \frac{1}{Z_q} \exp_q\left( -
\beta_q \mu_T \cosh y \right) \label{eq:formulaq}
\end{equation}
(where $\exp_q (...)$ is defined by eq. (\ref{eq:qexp}) and $Z_q$ is
given, as in (\ref{eq:Z}), by the normalization condition, $\int dy
p_q(y) = 1$). The characteristic feature of $p_q(y)$, as shown in
Fig. {\ref{fig:qrap}, is that it enhances (depletes) the  tails of the
distribution for $q>1$ ($q<1$), respectively (or, in other words, it
enhances or depletes the more or, respectively, less probable
events). Notice that in this case, differently than in
(\ref{eq:formula}), one has to be sure that $1-(1-q)\beta_q\mu_T\cosh
y \ge 0$, which imposes an  additional condition on the allowed phase
space for $q<1$ and $\beta_q>0$. In this case the strict correlation
between the shape of rapidity distribution and its height is relaxed
because both depend also on the new parameter $q$. This fact will be
important later on. The scaling-like formula (\ref{eq:formulas}) and
the approximate relation (\ref{eq:betaK}), this time between
$\beta_q$ and $\kappa_q$, are still valid, albeit this time only
approximately, i.e., for small values of $|q-1|$. On the other hand
$\beta_q >0$ for $N = \left(2\ln N_{max}\right)^q$, i.e., for larger
(smaller) multiplicities, depending whether $q>1$ ($q<1$).\\

\begin{figure}[h]
\begin{minipage}[t]{0.475\linewidth}
\centering
\includegraphics[height=7.cm,width=7.cm]{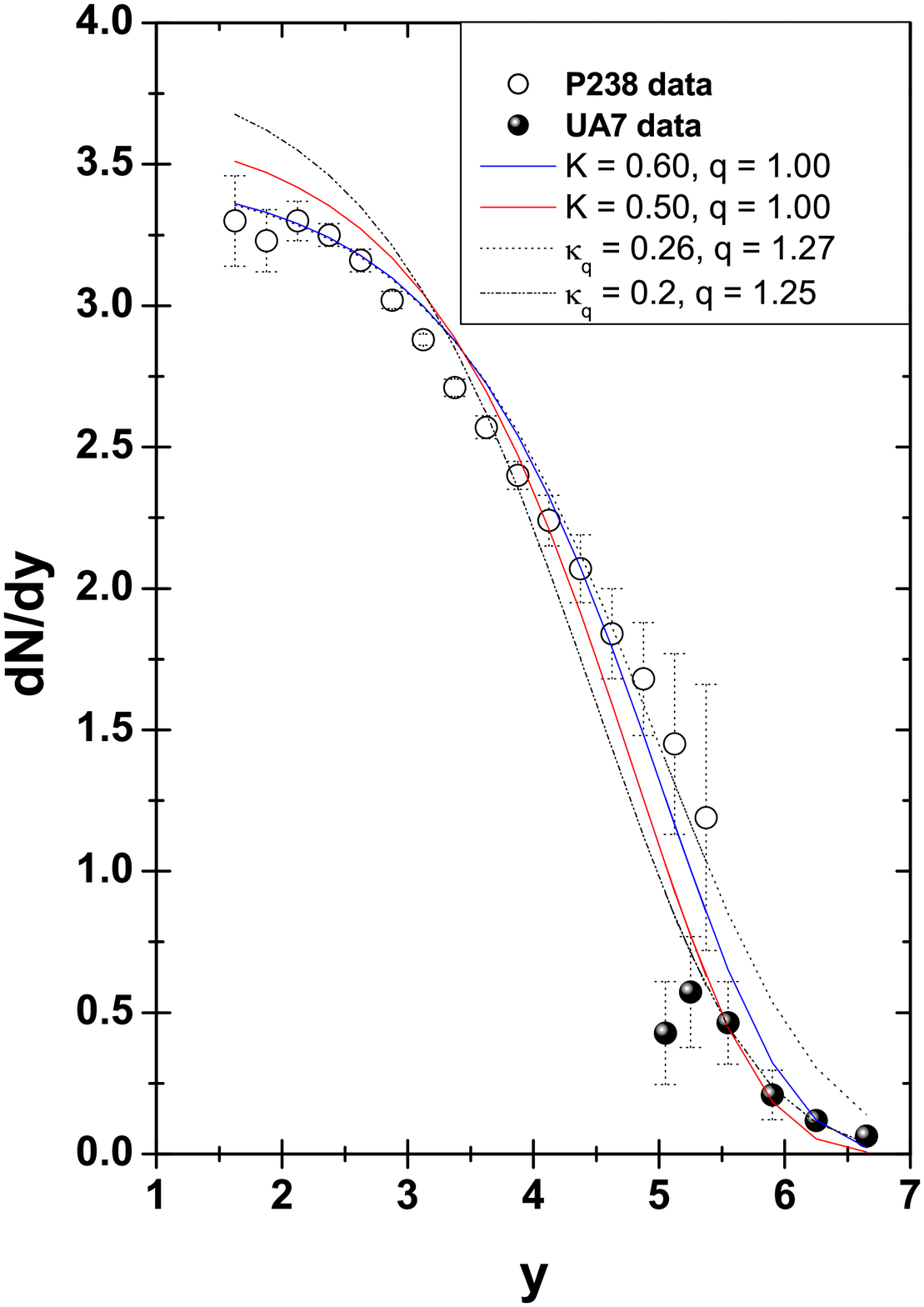}
\caption{Comparison of data for rapidity distributions at $\sqrt{s}=630$
GeV obtained by P238 \cite{P238} (open circles) and UA7 \cite{UA7}
collaborations (full circles) with predictions of our model for it
extensive ($q=1$) and nonextensive versions. Notice that one can fit
at the same time either P238 or UA7 data but not both together.}
\label{fig:630q}
\end{minipage}
\hspace{2mm}
\begin{minipage}[t]{0.475\linewidth}
\centering
\includegraphics[height=7.0cm,width=7.5cm]{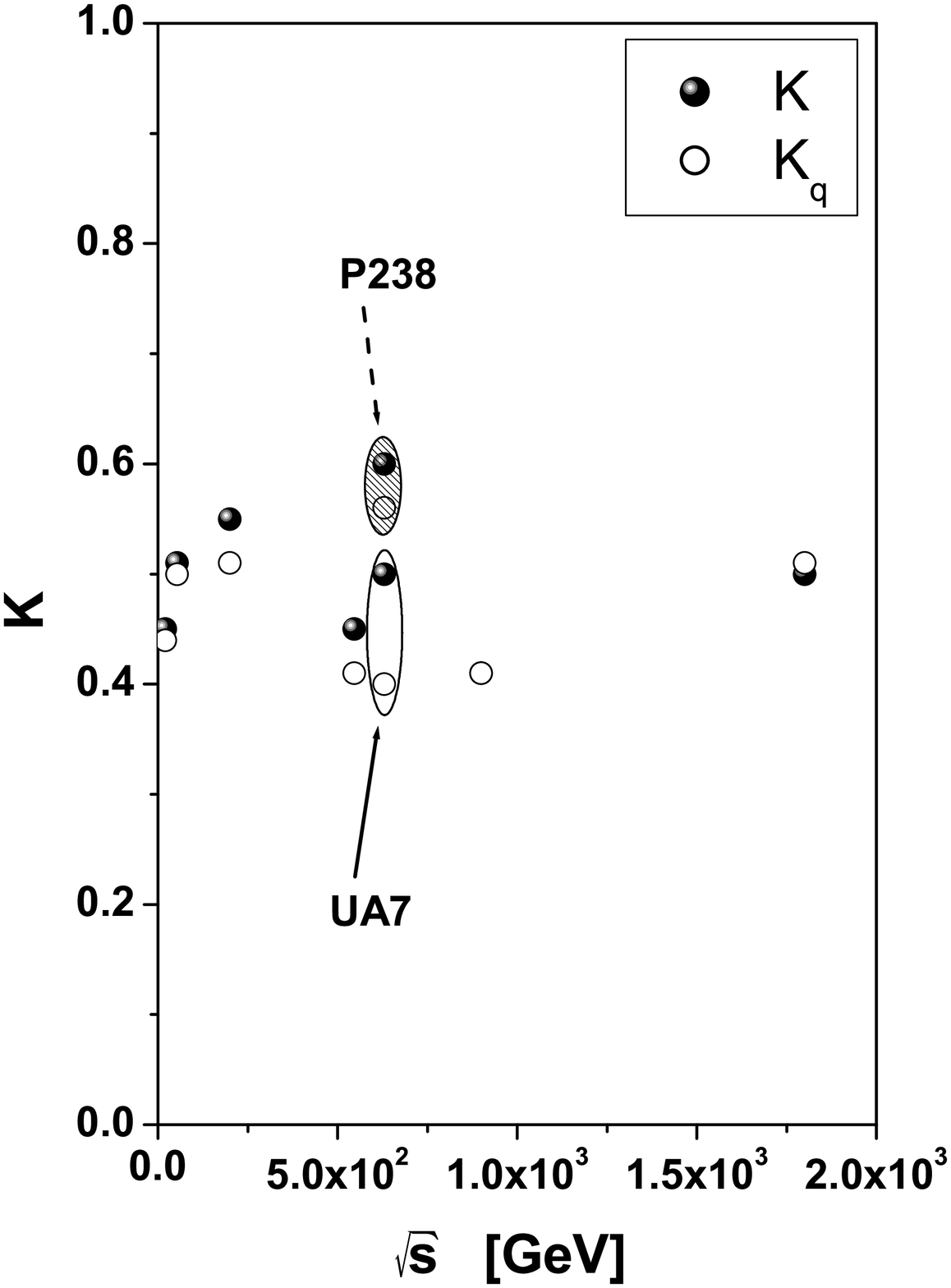}
\caption{Energy dependences of inelasticities obtained in extensive
($K$, see (\ref{eq:W})) and nonextensive ($K_q$, see
(\ref{eq:Koutq})) approaches (cf. also Table \ref{table:TI}). Notice
that results for P238 data do not follow the overall trend (see
text).}  
\label{fig:kvss}
\end{minipage}\hspace{2mm}
\end{figure}

The results for fits using extensive formula (\ref{eq:formula}) are
displayed in Fig. \ref{fig:fig1a} and those using its nonextensive
version given by eq. (\ref{eq:formulaq}) are shown in Fig.
\ref{fig:fig1b}. As one can see they are almost identical,
differences are showing up only for $20$ GeV where data develop a tail
which is most sensitive to the parameter $q$. The result for the
joint distributions of UA7 data from \cite{UA7} and P238 \cite{P238}
data at $630$ GeV are shown separately in Fig. \ref{fig:630q}. Both
sets of data are clearly incompatible in the sense that UA7 data
(which are for $\pi^0$'s and have been taken here in the same way as
in \cite{Ohsawa}) do not continue the trend shown by P238 data
(which are, as UA5 ones, for charged particles). Instead they seem to
continue the trend of the charged UA5 data, which is also clearly seen
from the values of the obtained parameters displayed in Table
\ref{table:TI}. Because of this fact we have fitted them separately.
The results for the respective inelasticity parameter $K$ and its
$q$-equivalent $\kappa_q$ for different energies are shown in Table
\ref{table:TI}. The estimated errors (the same for both approaches)
range from $\Delta K = 0.02$ for $20$ GeV (where the fitted
range in rapidity is biggest) to $\Delta K = 0.05$ for $1800$ GeV
(where the lack of measured tails prevents a better fit). These errors
should be kept in mind when looking at Fig. \ref{fig:kvss}, which
summarizes the obtained inelasticities of different types. These
inelasticities can be compared with inelasticities $K_{|y|\leq 4}$
defined by the formula      
\begin{equation}
K_{|y|\leq 4} = \frac{N}{\sqrt{s}} \int^{Y_m=4}_{Y_m=-4}\!dy\ p(y)
[\mu_T \cosh y] , \label{eq:Kleq4} 
\end{equation}
i.e., in the same way as in \cite{Chou}, namely, by integrating over
the same part of the phase space limited by $|Y_{m}|= 4$. Notice
that values of $K_{|y|\leq 4}$ are systematically smaller than the
corresponding values of $K$ and are in a visible way decreasing with
energy. The reason for such behaviour is that $K_{|y|\leq 4}$ counts
the fraction of energy in a fixed domain in the phase space given by
the condition that $|y|\leq 4$ whereas our $K$ gives the energy used
for production of particles in the whole kinematically allowed region
$|y|\leq Y_m$ as defined by eq. (\ref{eq:Ymax}). Table \ref{table:TI}
contains also the corresponding values of the "partition temperature"
$T=\frac{1}{\beta}$ obtained from the eq. (\ref{eq:W}).\\ 

\begin{table}[h]
\begin{center}
\begin{tabular}{|c|c|c||c|c|c||c|c|c|c|c|}
\hline
        &  &    &      &      &      &      &      &    && \\
$\sqrt{s}$ & $\langle n_{ch}\rangle$ & $\langle p_T \rangle$ & $K$  &
$T$ & $K_{|y|\leq4}$ & $q$ & $\kappa_q$ & $\tau_q$ & $K_q$ &
$K_q^{(|y|\leq4)}$ \\    
 (GeV)  & & (GeV) &      & (GeV) &      &      &     & (GeV) &     &\\ 
        & &     &       &       &      &      &      &      &      & \\ 
\hline
\hline                 
        &  &      &        &         &        &        &        &
 & &  \\
~~20   &  $7.7$ & 0.30 & $0.45$ & $1.76$ & $0.53$  & $1.05$ & $0.40$ &  $2.07$ &
 $0.44$ & $0.60$    \\
        &  &      &        &         &        &        &        &
 & &  \\
~~53    & $13.0$ & 0.34 & $0.51$ &  $3.53$ & $0.50$ & $1.13$ & $0.38$ & $4.06$ &
 $0.50$ &  $0.57$  \\ 
        &  &      &         &         &        &        &        &
 & &  \\ 
~200    & $21.4$ & 0.40 & $0.55$ & $12.12$ & $0.37$ & $1.20$ & $0.30$ &  $11.74$ &
 $0.51$ &  $0.37$  \\
        & &       &         &         &        &        &        &
 & &  \\ 
~540    & $29.1$ & 0.45 & $0.45$ & $22.38$ & $0.22$ & $1.26$ & $0.20$ & $20.39$ &
 $0.41$ &  $0.22$  \\
        & &       &        &         &        &        &        &
 & &  \\ 
$^{(a)}$630 & $31.0$ & 0.45 & $0.60$ & $36.29$ & $0.22$ & $1.27$ & $0.26$ &  $35.51$ &
 $0.56$ &  $0.22$  \\
        &  &      &        &         &        &        &        &
 & &  \\        
$^{(b)}$630 & $31.0$ & 0.45 & $0.50$ & $28.90$ & $0.21$ & $1.25$ & $0.20$ &  $21.22$ &
 $0.40$ &  $0.21$  \\
        & &       &        &         &        &        &        &
 & &  \\ 
 ~900   & $34.6$ & 0.48 & $0.41$ & $29.47$ & $0.17$ & $1.29$ & $0.18$ & $30.79$ &
 $0.41$ &  $0.17$  \\
        & &       &        &         &        &        &        &
 & &  \\ 
 1800   & $46.4$ & 0.50 & $0.50$ & $55.69$ & $0.13$ & $1.33$ & $0.19$ & $62.57$ &
 $0.51$ &  $0.13$   \\
        & &       &        &         &        &        &        &
 & &  \\ 
\hline        
\end{tabular}
\caption{The results for extensive (eqs. (\ref{eq:formula})) and
nonextensive (eq. (\ref{eq:formulaq})) approaches applied to $pp$
\cite{NucData2} and $p\bar{p}$ data \cite{UA5,Tevatron,P238,UA7} on
rapidity distributions. The first three columns summarize our input
information: energy $\sqrt{s}$, total charge multiplicities $\langle
n_{ch}\rangle$ (estimated as in \cite{N}) and $\langle p_T\rangle$
(estimated as in \cite{NB}). Presented are extensive inelasticity $K$
and parameters $\kappa_q$ and $q$ out of which the nonextensive 
inelasticity $K_q$ is calculated by means of (\ref{eq:Koutq})). For
completeness the corresponding "partition temperatures" $T=1/\beta$
and $\tau_q = 1/\beta_q$ are also listed as well as the corresponding
inelasticities obtained for the limited portion of the phase space:
$K_{|y|\leq 4}$ as given by eq. (\ref{eq:Kleq4}) and
$K_q^{(|y|\leq4)}$ as defined by (\ref{eq:Kqleq4}). For $\sqrt{s} =
630$ GeV we display separately results from fitting $(a)$ P238 data
\cite{P238} and $(b)$ UA7 data \cite{UA7}.}    
\label{table:TI}
\end{center}
\end{table}

In what concerns the nonextensive approach one must realize that
parameter $\kappa_q$ occurring in (\ref{eq:Wq}) is not the
inelasticity in the same sense as $K$ from the extensive approach,
cf. eq. (\ref{eq:W}). The reason is simple (and shown in best way in
\cite{Beck} where Hagedorn statistical model of multiparticle
production \cite{H} has been extended to $q$-statistics). Namely,
because $q$ summarizes all kinds of correlations and/or fluctuations
present in the system (and make it nonextensive) the energies per
particle present on the rhs of eqs. (\ref{eq:W}) and (\ref{eq:Wq})
also contain in the nonextensive case  contributions from these
correlations or fluctuations, i.e., a kind of effective interaction
characterized by $|q-1|$ \cite{Beck,WWq}. The enhanced (depleted) for
$q>1$ ($q<1$) tails of $p_q(y)$ observed in Fig. \ref{fig:qrap} say
that more (less) particles are sent there towards the end of the phase
space, respectively. This fact must then be compensated by the
appropriate choice of the value of "$q$-inelasticity" parameter
$\kappa_q$ in (\ref{eq:Wq}), which fixes the energy $W_q$ in this
case. It should be also added at this point that when one is allowing
the whole energy $\sqrt{s}$ to be used for the production of
secondaries and uses nonextensive version of information theory to
find $\frac{dN}{dy}$ then one finds $q<1$ \cite{NC}, not $q>1$ as
here. This is, however, expected because of the already mentioned
fact that the $q<1$ case enhances frequent events whereas $q>1$ the rare
ones. When considering the whole phase space it comprises all
produced particles, which are located predominantly only in part of
it. Therefore we have to enhance those frequent events by using
$q<1$. This choice, as was mentioned above, limits also the allowed
phase space. On the other hand, when $K$ is accounted for (as in the
present case), the allowed phase space is  essentially correctly
described by $K\cdot \sqrt{s}$ and one has only to enhance the rare
events when particles (because of fluctuations) "leak out" of it,
what results in $q>1$.\\   

\begin{figure}[h]
\begin{minipage}[t]{0.475\linewidth}
\centering
\includegraphics[height=7.0cm,width=7.0cm]{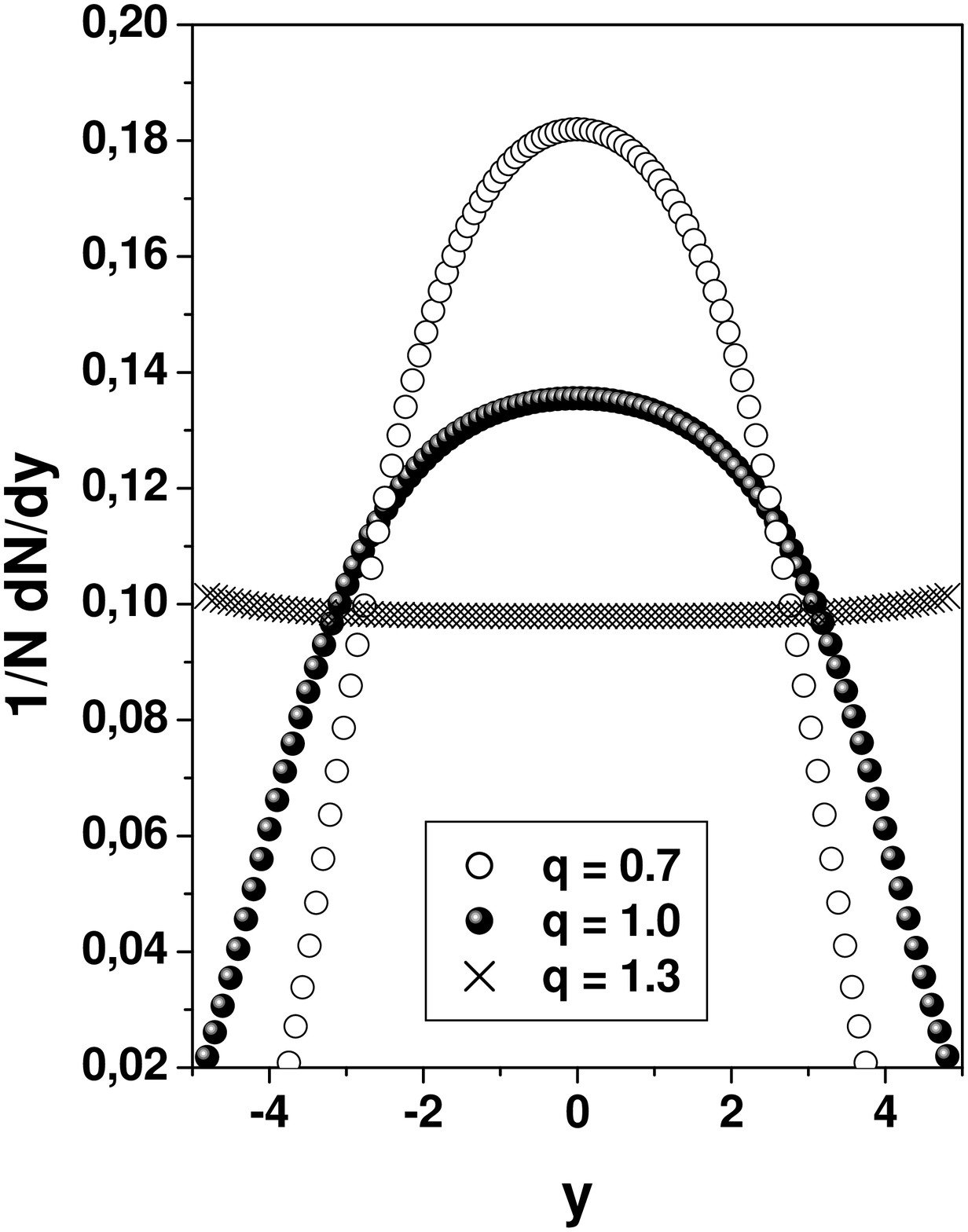}
\caption{The examples of $p_q(y)$ as given by eq.(\ref{eq:formulaq})
for $q=0.7$ and $q=1.3$ compared with $p(y)$ for $q=1$ as given by
eq.(\ref{eq:formula}) for one-dimensional hadronization of mass
$M=100$ GeV into $N=20$ secondaries.} 
\label{fig:qrap}
\end{minipage}\hspace{2mm}
\hfill
\begin{minipage}[t]{0.475\linewidth}
\centering
\includegraphics[height=6.7cm,width=7.0cm]{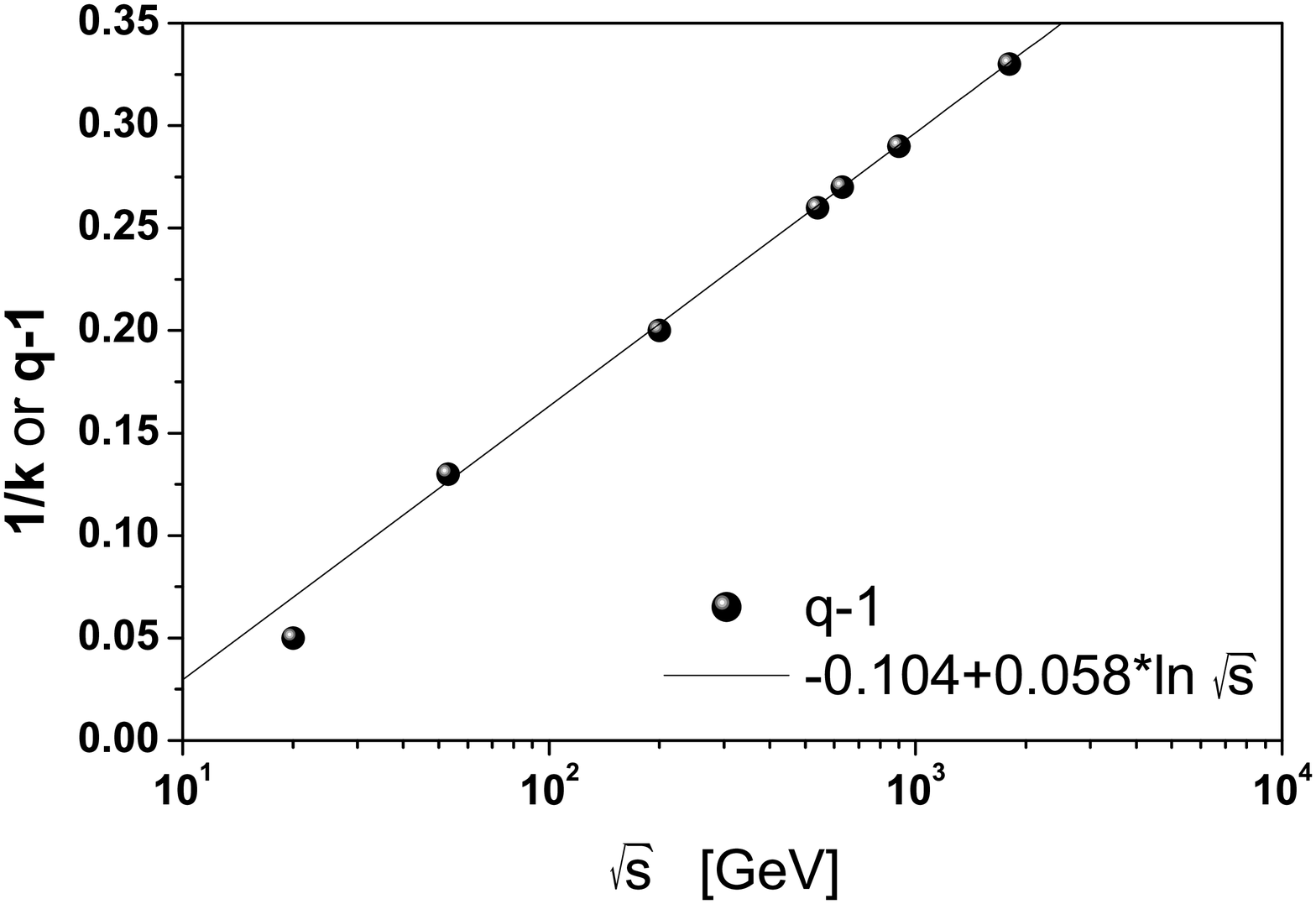}
\caption{The values of the nonextensivity parameter $q$ obtained in
fits shown in Fig. \ref{fig:fig1b} (and listed in Table
\ref{table:TI}) compared with the values of the parameter $k$ of
Negative Binomial distribution fit to the corresponding multiplicity
distributions as given in \protect\cite{NB}, see eq.(\ref{eq:1ok}).}. 
\label{fig:kvsq}
\end{minipage}
\end{figure}

In order to get a nonextensive version of inelasticity, i.e., $K_q$, let us
first observe that, according to (\ref{eq:W}), inelasticity $K$ can
be expressed by the  mean energy per particle, $\langle E\rangle$,
therefore, in the nonextensive case one can write accordingly 
\begin{eqnarray}
K_q &=& \frac{N}{\sqrt{s}}\langle E\rangle_q\, =\,
\frac{N}{\sqrt{s}}\int^{Y_m}_{Y_m}\!dy\ p_q(y) [\mu_T \cosh y] ,
\label{eq:Koutq}  \\
&\approx & \frac{\kappa_q}{3-2q} .\nonumber
\end{eqnarray}
(with $Y_m$ provided by eq. (\ref{eq:Ymax})). The approximate
relation of $K_q$ with the  parameters $\kappa_q$ and $q$ in nonextensive
formula (\ref{eq:formulaq}) arises when one estimates $\langle
E\rangle_q$ for $|Y_m| \rightarrow \infty$ and uses the nonextensive
version of relation (\ref{eq:betaK}). As it can be seen from Table
\ref{table:TI},  nonextensive inelasticity $K_q$ defined this way
agrees reasonably well with the extensive inelasticity $K$. Notice
that $K_q^{(|y|\leq4)}$, defined as  
\begin{equation}
K_q^{(|y|\leq 4)} = \frac{N}{\sqrt{s}} \int^{Y_m=4}_{Y_m=-4}\!dy\
p_q(y) [\mu_T \cosh y] , \label{eq:Kqleq4} 
\end{equation}
is essentially identical with $K_{|y|\leq4}$ discussed above (and for
the same reason). These results indicate that the true equivalent of
inelasticity $K$ in nonextensive approach is $K_q$, at least in the
sense used in cosmic ray research, namely that it defines the part of
the initial energy taken away by leading particles: $K_{elastic} = 1
- K$ should be replaced by: $K_{elastic} = 1 - K_q$. They are
displayed explicitly in  Fig. \ref{fig:kvss}. Notice that P238 data
\cite{P238} at $\sqrt{s}=630$ GeV clearly do not follow the trend
presented by the UA5, UA7 and Tevatron data \cite{UA5,Tevatron}. When
neglecting this point the overall tendency is that inelasticity is
essentially constant with energy and equal to $K\simeq 0.5$, which agrees
with first estimates made in \cite{Cocconi,CAS3,Inel} and with
first experimental estimates based on the analysis of the leading
particle effect provided in \cite{Basile}.\\    

\begin{figure}[h]
\begin{center}
  \begin{minipage}[ht]{6cm}
    \centerline{\epsfig{file=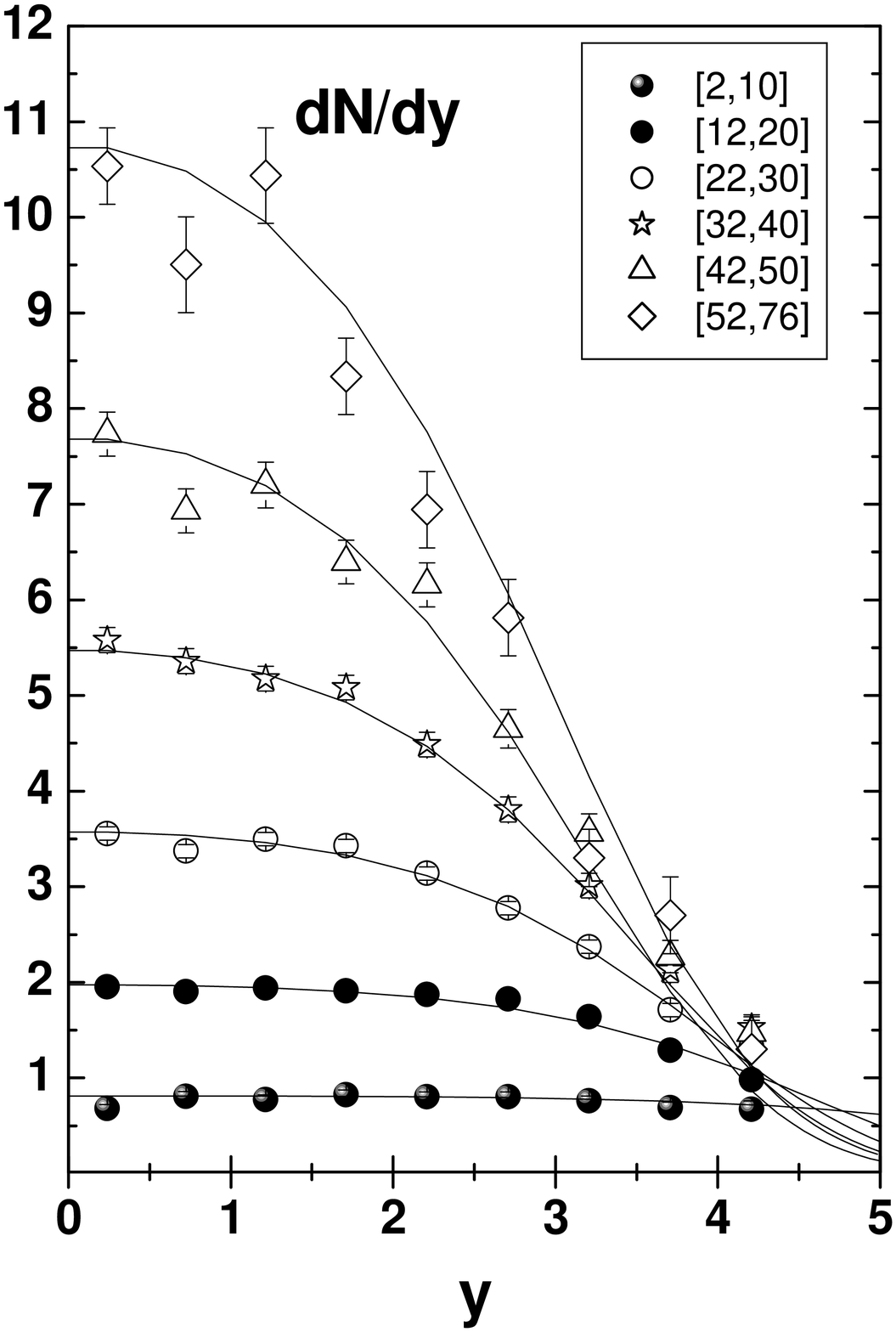, width=62mm}}
  \end{minipage}
\hspace{2cm}
  \begin{minipage}[ht]{6cm}
    \centerline{\epsfig{file=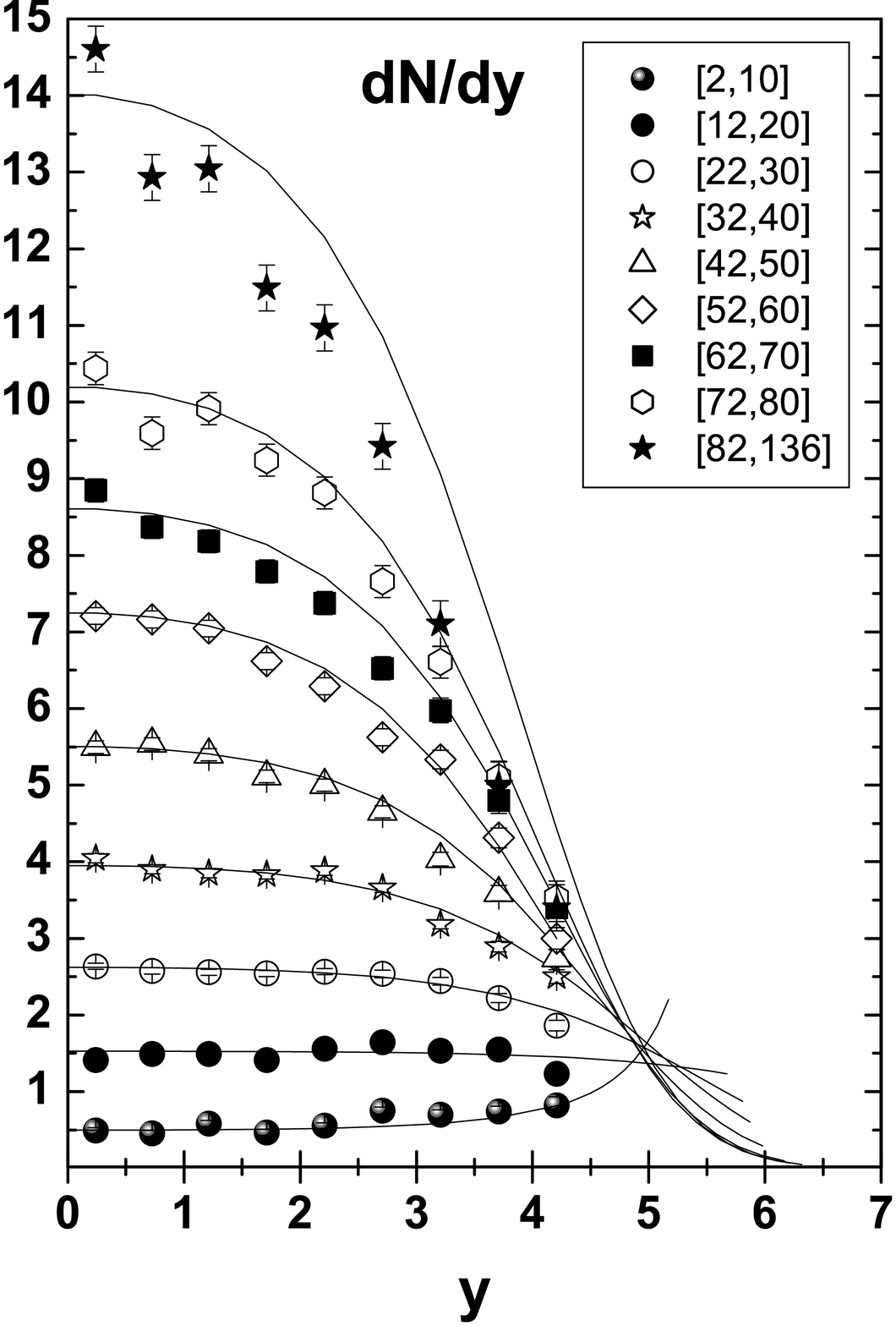, width=58mm}}
  \end{minipage}
\end{center}
\caption{Fits to $\frac{dN}{dy}$ for different rapidity bins for
$\sqrt{s}=200$ and $900$ GeV (left and right panels, respectively) by
means of eq.(\ref{eq:formulaq}).}
\label{fig:yn}
\end{figure}

A comment on the possible physical meaning of the parameter $q$
obtained from our fits and listed in Table \ref{table:TI} is in order
here. As we said before, in general, the nonextensivity parameter $q$
summarizes the action of several factors, each of which leads to a 
deviation from the simple form of the extensive Boltzmann-Gibbs
statistics, or Shannon entropy defined by eq.(\ref{eq:Shannon}).
Among them are the possible intrinsic fluctuations existing in the
hadronizing system \cite{WWq}. Notice that in the case considered
here we have not accounted explicitly for the fact that each event
has its own multiplicity $N$ but we have used only its mean value,
$\langle N\rangle$, as given by experiment where $\langle N\rangle =
\sum N P(N)$ with $P(N)$ being the multiplicity distribution.
Actually, we have used only its charged part, $\langle n_{ch}\rangle$,
assuming that $N = \frac{3}{2}\langle n_{ch}\rangle$, i.e.,
neglecting in addition also possible fluctuations between the number
of charged and neutral secondaries. Experimentally 
it is known that $P(n_{ch})$ is adequately described by the so called
Negative Binomial distribution (NBD) \cite{NB}, which depends on two
parameters: the mean multiplicity $\langle n_{ch}\rangle$ and
the parameter $k$ ($k\ge 1$) affecting its width,
\begin{equation}
\frac{1}{k}\, =\, \frac{\sigma^2(n_{ch})}{\langle n_{ch}\rangle^2}\, -\,
\frac{1}{\langle n_{ch}\rangle}. \label{eq:k} 
\end{equation}
For $k\rightarrow 1$ NB approaches a geometrical distribution whereas
for $k^{-1}\rightarrow 0$ it approaches a Poissonian distribution.
In general it is found \cite{NB} that 
\begin{equation}
\frac{1}{k} = -0.104 + 0.058\cdot \ln \sqrt{s}. \label{eq:1ok}
\end{equation}
Following the  ideas expressed in \cite{WWq} we would like to draw 
attention to the fact that the value of $k^{-1}$ may be also
understood as the measure of fluctuations of the mean multiplicity
(cf. also \cite{ISVHECRI}). When there are only statistical
fluctuations in the hadronizing system one should expect the
Poissonian form of the corresponding multiplicity distributions. 
The existence of intrinsic
(dynamical) fluctuations would mean that one allows the mean
multiplicity $\bar{n}$ to fluctuate. In the case when such
fluctuations are given by a gamma distribution with normalized variance
$D(\bar{n})$ then, as a result, one obtains the Negative Binomial
multiplicity distribution with 
\begin{equation}
\frac{1}{k}\, =\, D(\bar{n})\, =\,
\frac{\sigma^2\left(\bar{n}\right)}{\langle \bar{n}\rangle^2} . \label{eq:D}
\end{equation}
That is because in this case (see also \cite{Shih}):
\begin{equation}
P(n)\, =\, \int_0^{\infty} d\bar{n} \frac{e^{-\bar{n}}\bar{n}^n}{n!}\cdot 
         \frac{\gamma^k \bar{n}^{k-1} e^{-\gamma \bar{n}}}{\Gamma (k)} =
   \frac{\Gamma(k+n)}{\Gamma (1+n) \Gamma (k)}\cdot 
   \frac{\gamma^k}{(\gamma +1)^{k+n}} \label{eq:PNBD}
\end{equation}   
where $\gamma = \frac{k}{\langle \bar{n}\rangle}$. Assuming now that
these fluctuations contribute to nonextensivity defined by the
parameter $q$, i.e., that $D(\bar{n}) = q-1$ \cite{WWq} one should
expect that \cite{AK}   
\begin{equation}
q\, =\, 1\, +\, \frac{1}{k}. \label{eq:qk}
\end{equation}
As can be seen in Fig. \ref{fig:kvsq} {\it this is precisely the
case}. Namely, fluctuations existing in experimental data for the
rapidity distributions, $dN/dy$ \cite{UA5,Tevatron,NucData2}, and
disclosed by fits using the nonextensive form (\ref{eq:formulaq}) 
follow (except for the lowest energy point at $20$ GeV) the  pattern of
fluctuations seen in data for multiplicity distributions $P(N)$ and
summarized by the parameter $k$ of NBD \cite{NB}. This means then that
these data contain no more information than used here, namely
existence of limited $p_T$, inelasticity $K$ and fluctuations as
given by $q>1$ or $k^{-1}>0$.\\  

\begin{table}[h]
\begin{center}
\begin{tabular}{|c||c|c|c|c||c|c|c|c||}
\hline
           &\multicolumn{4}{c||}{} & \multicolumn{4}{c||}{} \\
           &\multicolumn{4}{c||}{$\sqrt{s} = 200$ GeV} & 
            \multicolumn{4}{c||}{$\sqrt{s} = 900$ GeV} \\           
$\Delta N$ &\multicolumn{4}{c||}{} & \multicolumn{4}{c||}{} \\            
\cline{2-9}
           &       &       &       &               &       &        &
       &               \\ 
           & $q$   & $\kappa_q$ & $\tau_q$ & $K_q$ & $q$   
           & $\kappa_q$  & $\tau_q$  & $K_q$ \\ 
           &       &       & (GeV) &          &        &
( GeV) &         &      \\ 
\hline                 
           &       &       &       &               &       &        &
       &             \\ 

 $[2,10]$    & 1.001 & 0.35 & 114.00 & 0.35 & 1.25 & 0.10 & -34.20 &
 0.16 \\ 
       &       &      &        &      &      &      &        &
     \\
 $[12,20]$   & 1.10  & 0.41 &  20.99 & 0.54 & 1.26 & 0.17 & 316.55 &
 0.33 \\
       &       &      &        &      &      &      &        &
     \\
 $[22,30]$   & 1.14  & 0.44 &  11.00 & 0.64 & 1.27 & 0.20 &  62.63 &
 0.42 \\
       &       &       &       &      &      &      &        &
     \\
 $[32,40]$   & 1.17  & 0.45  &  7.31 & 0.72 & 1.27 & 0.22 &  34.98 &
 0.48 \\
       &       &       &       &      &      &      &        &
     \\
 $[42,50]$   & 1.18  & 0.46  &  5.10 & 0.75 & 1.23 & 0.25 &  22.74 &
 0.50 \\
       &       &       &       &      &      &      &        &
     \\
 $[52,N_1]$  & 1.25  & 0.50  &  4.39 & 0.98 & 1.21 & 0.26 &  16.28 &
 0.49 \\
       &       &       &       &      &      &      &        &
     \\
 $[62,70]$   &       &       &        &    & 1.20 & 0.31 &  15.79 &
 0.57 \\
       &       &       &       &      &      &      &        &
     \\
 $[72,80]$   &       &       &        &   & 1.20 & 0.33 &  14.02 &
 0.61 \\
       &       &       &       &      &      &      &        &
     \\
 $[82,136]$  &       &       &        &   & 1.22 & 0.38 &  12.00 &
 0.74 \\
       &       &       &       &      &      &      &        &
     \\
\hline 
\end{tabular}
\caption{Results for the same parameter $q$, $\kappa_q$, $\tau_q$ and
$K_q$ as shown and defined in Table \ref{table:TI} but now obtained
for data taken in the restricted multiplicity bins $\Delta N$
\protect\cite{DeltaN} and shown in Fig. \ref{fig:yn}. Here: $N_1=76$
for $\sqrt{s} = 200$ and $60$ for $900$ GeV.} 
\label{table:TII}
\end{center}
\end{table}

\begin{figure}[h]
\centering
\includegraphics[height=7.0cm,width=10.0cm]{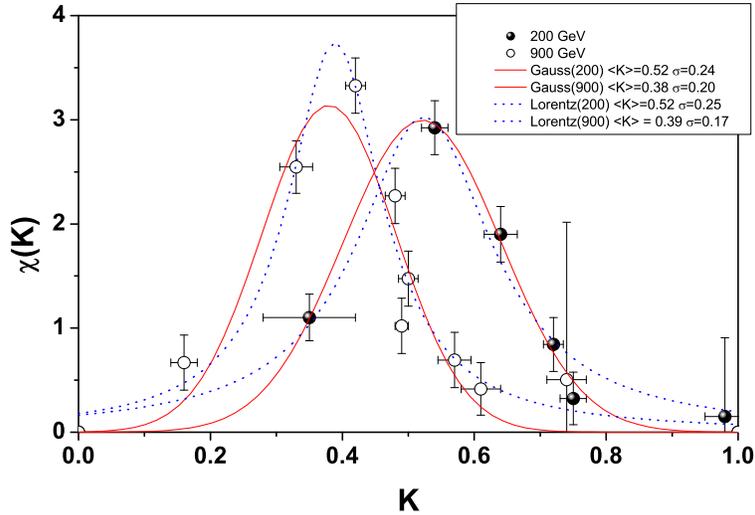}
\caption{Inelasticity distributions $\chi(K=K_q)$ (normalized
to unity) obtained from \cite{DeltaN} data for $\sqrt{s} = 200$ GeV
and $\sqrt{s} = 900$ GeV (see text for details). $K_q$ is
estimated from the $q$ and $\kappa_q$ parameters in Table
\ref{table:TII} by using eq.(\ref{eq:Koutq}). To guide the eyes
obtained results were fitted by simple gaussian and lorentzian
formulas ($\chi(K) \simeq \exp \left[ -(K-\langle
K\rangle)/(2a^2)\right]$ and $\chi(K) \simeq  \sigma/[4(K-\langle
K\rangle)^2 + \sigma^2]$, respectively, with $\sigma$ obtained for
$K\in (0,1)$). Errors are estimated from the widths of the
bins $\Delta N$ and the shapes and errors of $P(N)$ used in obtaining
$\chi(K)$ \cite{NB}.}    
\label{fig:khi2}
\end{figure}

It is interesting to notice that whereas  data on rapidity
distributions \cite{UA5,Tevatron,P238,UA7} could be fitted {\it both}
by the extensive (\ref{eq:formula}) and nonextensive (\ref{eq:formulaq}) 
distributions, similar data for rapidity distributions measured in
restricted intervals of the multiplicity, $\Delta N$ \cite{DeltaN},
can be fitted {\it only} by means of the nonextensive $p_q(y)$ as
given by eq.(\ref{eq:formulaq}) \cite{FOOT3}. The extensive approach
with maximally correlated shapes and heights of distributions, as
discussed above, is clearly too restrictive. Only relaxing this
correlation by introducing parameter $q$ (i.e., by using nonextensive
version of the information theory approach) allows for adequate fits
to be performed, see Fig. \ref{fig:yn} and Table \ref{table:TII}.
Notice that partial inelasticities of all kinds are clearly
correlated with the multiplicity bins, the higher the multiplicity the
bigger is the corresponding inelasticity. The same kind of
correlations are observed at $200$ GeV between multiplicity and $q$,
which increases with multiplicity. However, at $900$ GeV $q$ remains
essentially constant. Following discussion in the previous paragraph
one expects that it means an  increase of the corresponding fluctuations.
The question, however, remains, in which variable? We 
argue that the fluctuating variable in this case is inelasticity itself.
The point is that particles filling a given interval of multiplicity
$\Delta N$ can be produced in events with different, i.e., {\it
fluctuating} values of $K$. As before, this fact would then lead to
the apparent nonextensivity visualized by $q>1$ and measuring also
the strength of such fluctuations represented by the variance, 
\begin{equation}
\sigma^2(K)\, =\, \langle K\rangle^2 (q-1). \label{eq:sigmaK}
\end{equation}
However, in this case we do not have any independent estimation of
$\sigma(K)$, therefore we could not exclude the action of some other, so
far not yet disclosed, factors and  propose  the equivalent of eq.
(\ref{eq:qk}) for this case. On the other hand (\ref{eq:sigmaK}) could
be used for estimation of the uncertainty in $K$ once its mean value
and the nonextensivity parameter are known. For example, taking from
Table \ref{table:TI} the corresponding values of $\langle
K(s)\rangle$ and $q(s)$ one can estimate that $\langle
K_q(200)\rangle =0.50 \pm 0.23$ and $\langle K_q(900)\rangle =
0.41\pm 0.22$.\\ 

The results for partial inelasticities obtained from fits shown in
Fig. \ref{fig:yn} allow us to calculate the corresponding
inelasticities $K_q$ (by using (\ref{eq:Koutq})), cf. Table
\ref{table:TII}. These in turn, with the help of experimentally
measured  multiplicity distributions $P(n_{ch})$ (taken in this case
from \cite{P(N)}) allow us to obtain, {\it for the first time}, the
(normalized) inelasticity distribution $\chi(K=K_q)$
presented in Fig. \ref{fig:khi2} \cite{FOOTalfa}. This is one of the
most important results which could be obtained only by using
information theory approach in its nonextensive version. The gaussian
and lorentzian fits shown here resemble very much the form of
$\chi(K)$ obtained in the so called Interacting Gluon Model of high
energy processes developed and studied in \cite{INEL,Fernando,LEADING}.\\

\section{SUMMARY AND CONCLUSIONS}

Using methods of information theory, both in its extensive and
nonextensive versions, we have analysed $p\bar{p}$ collider data
\cite{UA5,Tevatron,P238,UA7,DeltaN} and $pp$ fixed target data
\cite{NucData2} on multiparticle production. Our investigation was
aimed at the phenomenological, maximally model independent
description, which would eventually result in estimations of
inelasticities for these reactions, their energy dependence and,
whenether possible, also in the inelasticity distributions $\chi(K)$.
The information theory approach used by us leads to a comfortable
situation where the only fitted parameter is either the inelasticity
$K$ (when using extensive approach) or parameters $q$ and $\kappa_q$
(in its nonextensive counterpart) out of which one can reconstruct
the inelasticity $K_q$ or $K_q^{(|y|\leq 4)}$. It turned out, that data for
rapidity distributions obtained for the mean multiplicities can be
fitted using both approaches. In this case the nonextensivity
parameter $q$ obtained from fitting the {\it rapidity} distributions 
$dN/dy$ is practically {\it identical} to the parameter $q$,
according to \cite{WWq}, responsible for dynamical fluctuations
existing in hadronizing systems and showing up in the characteristic
Negative Binomial form of the measured {\it multiplicity}
distributions $P(N)$. This  means therefore that in collider data for
$p\bar{p}$ collisions and fixed target $pp$ data analysed in  this way,
there is no additional information to that used here. On the other
hand rapidity distributions for fixed multiplicity intervals 
$\Delta N$ \cite{DeltaN} can be described only by nonextensive
approach and we argue that in this case $q$ reflects fluctuations in
the inelasticity itself. These data were therefore used to estimate,
for the first time at this energies ($200$ and $900$ GeV), the
inelasticity distributions $\chi(K)$, cf. Fig. \ref{fig:khi2}.\\ 

It is particularly interesting and worth of stressing here, that,  
formulas obtained by means of information theory  are apparently {\it
identical} with the corresponding equations of statistical models
used to describe multiparticle production processes \cite{Bec}. The
point is, however, that - as was already stressed in appropriate
cases before - the "partition temperature" $T$ and the normalization
$Z$ are in our case {\it not free parameters} anymore. The only
freedom (in our case it was the choice of inelasticity $K$) is in
providing the corresponding constraint equations, which should
summarize our knowledge about the reaction under consideration. Once
they are fixed,  the other quantities (in particular $T$) follow. This
constrains seriously such approach and therefore in cases  where it fails
one can either add new constraints or  include some interactions by
changing the very definition of how to measure the available
information. The Tsallis entropy used here \cite{T,WWq,CBeck,Abe,Beck} is
but only one example of what is possible, other definitions of
information are also possible albeit not yet used in such
circumstances \cite{T,Frieden}.\\

%\newpage


\begin{thebibliography}{9}
\bibitem{FOOTA} It could be argued that by allowing the growth of
$\langle p_T\rangle$ with $\sqrt{s}$ we also include, at least to some
extent, the growing with energy influence of hard collisions
visualized as occurrence of the so called gluonic mini-jets, cf. 
\cite{Fernando}.
\bibitem{FOOT3} Actually, only data on $\Delta N$ for $\sqrt{s} =
200$ and $900$ Gev will be used here because we have found that data
for $540$ GeV (used previously in \cite{Chou}), are incompatible with
the corresponding data for the mean multiplicities $<N>$.
\end{thebibliography}
\end{document}